\documentstyle[NATO,psfig]{crckapb} 

\begin{opening}

\title{Scheduled Discoveries of 7+ High-Redshift Supernovae: \protect\\
First Cosmology Results and Bounds on $\lowercase{q}_0$
}
\subtitle{The Supernova Cosmology Project: I}

\author{S. Perlmutter$^1$}
\author{S. Deustua}
\author{S. Gabi}
\author{G. Goldhaber}
\author{D. Groom}
\author{I. Hook}
\author{A. Kim}
\author{M. Kim}
\author{J. Lee}
\author{R. Pain$^2$}
\author{C. Pennypacker}
\author{I. Small}
\institute{E. O. Lawrence Berkeley National Laboratory \& Center
for Particle Astrophysics, 
University of California, Berkeley}
\author{A. Goobar}
\institute{University of Stockholm}
\author{R. Ellis}
\author{R. McMahon}
\institute{Institute of Astronomy, Cambridge University}
\author{B. Boyle}
\author{P. Bunclark}
\author{D. Carter}
\author{K. Glazebrook$^3$}
\author{M. Irwin}
\institute{Royal Greenwich Observatory}
\author{H. Newberg}
\institute{Fermi National Accelerator Laboratory}
\author{A. V. Filippenko}
\author{T. Matheson}
\institute{University of California, Berkeley}
\author{M. Dopita}
\author{J. Mould}
\institute{MSSSO, Australian National University}
\author{W. Couch}
\institute{University of New South Wales}

\end{opening}

\runningtitle{HIGH REDSHIFT SUPERNOVAE}

\begin{document}

\footnotetext[1]{Presented by S. Perlmutter, e-mail address:{\it saul@LBL.gov}}
\footnotetext[2]{Current address:  CNRS-IN2P3, University of Paris}
\footnotetext[3]{Current address: Anglo-Australian Observatory}

\begin{abstract}

Our search for high-redshift Type Ia supernovae discovered, in its first years,
a sample of seven
supernovae.  Using a ``batch'' search strategy, almost all were discovered 
before maximum light and were observed over the peak of their light curves.
The spectra and light curves indicate that almost all were Type Ia supernovae 
at redshifts $z = 0.35$ -- 0.5.  These high-redshift supernovae can provide a 
distance indicator and ``standard clock''  to study the cosmological parameters 
$q_0$, $\Lambda$, $\Omega_0$, and $H_0$.  This presentation 
and the following presentations of Kim {\it et al.} (1996), 
Goldhaber {\it et al.} (1996),  and Pain {\it et al.} (1996)
will discuss observation strategies and rates, analysis and calibration issues,
the sources of measurement uncertainty, 
and the cosmological implications, including 
bounds on $q_0$, of
these first high-redshift supernovae from our ongoing search.

\end{abstract}
\section{Introduction}

Since the mid 1980's, soon after the identification of the 
supernova subclassifications Ia and Ib were recognized, Type Ia 
supernovae (SNe Ia) have appeared likely to be homogeneous 
enough that they could be used for cosmological 
measurements.  At the time, it appeared that they could 
be used to determine $H_0$, if their absolute 
magnitude at peak could be measured, i.e., if some SN Ia's 
distance could be calibrated.  They could also be used to
determine 
$q_0$ from the {\em apparent} magnitudes and redshifts of 
nearby and high redshift supernovae, if high redshift SNe Ia 
could be found.  Many of the presentations at this meeting 
have demonstrated significant advances in recent years in  our 
understanding of the SN Ia homogeneity (and/or 
luminosity calibration), 
the distances to SNe Ia, and measurements 
of $H_0$.  We will here discuss the search for high redshift 
supernovae and the measurement of $q_0$.

\section{High Redshift Supernovae Search Strategy}

We began our High Redshift Supernova Search soon after the 
completion of a labor-intensive search at the Danish 1.5-m in Chile, 
which in two years of searching had discovered one SN Ia  at 
$z = 0.31$, approximately 18 days past maximum light 
(N\mbox{\o}rgaard-Nielsen {\it et al.} 1989\nocite{no:Danish}).  
It was apparent that SNe Ia at 
high redshift are difficult to work with for at least three 
reasons: they are rare, they are rapid, and they are random.  
The estimates of SN Ia rates---a few per millennium per 
galaxy---are daunting, if one wants a statistically useful 
sample of supernovae.  Much of the interesting data must be 
obtained rapidly, since the supernova rises to maximum light 
within a few weeks and, at high redshifts, fades below the 
largest telescopes'  limits within a month or two.  
Furthermore, it is not possible to guarantee photometry and 
particularly spectroscopy of randomly occurring 
high-redshift supernovae, since the largest, most over-scheduled 
telescopes are needed to observe them.

Ideally one would like to {\em schedule} supernova 
explosions on demand, and then apply for the telescope time 
to study them, beginning at least a few days before maximum
light.

\begin{figure}
\psfig{figure=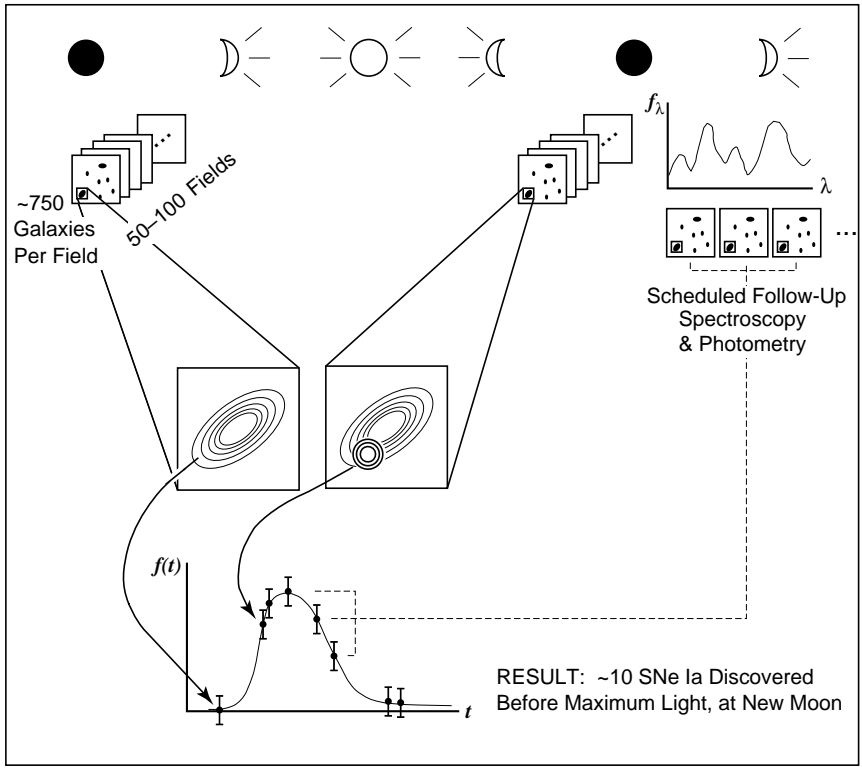,height=4in}
\caption[figurecaption]{
 Search strategy designed to discover batches of $\sim$10 high-redshift
supernovae on demand, just before new moon, and while the supernovae are
still brightening, i.e. before ``maximum light.''  The follow-up spectroscopy and
photometry can therefore be scheduled, and can follow the supernovae over
maximum light during dark time. }
  \label{strategy}
\end{figure}

To solve these problems, we developed a new search 
technique.  Figure 1 presents a schematic outline of the
strategy.  Just after a new moon, we observe many tens of 
high-galactic-latitude fields (including known high-redshift 
clusters when possible) on a 2.5- to 4-meter telescope.  
With a wide-field camera, each image contains hundreds of 
galaxies at redshifts 0.3 -- 0.6.  Just before the following 
new moon, we observe the same fields again.  We compare the
images, thus checking tens of thousands of high redshift 
galaxies (including those below our detection limit)  to find 
the ten
or so showing the new light of a supernova that was 
not there on the previous observation.  The supernovae 
generally do not have time to reach maximum light, with only 
2.5 to 3 weeks (or approximately 11 to 14 days in the 
supernova rest frame) between our after- and before-new-moon 
comparison images.  In order to begin the follow-up 
photometry and spectroscopy immediately, we have developed 
extensive software to make it possible to complete the 
analysis of all the images within hours of  the 
observations.

\begin{figure}
\psfig{figure=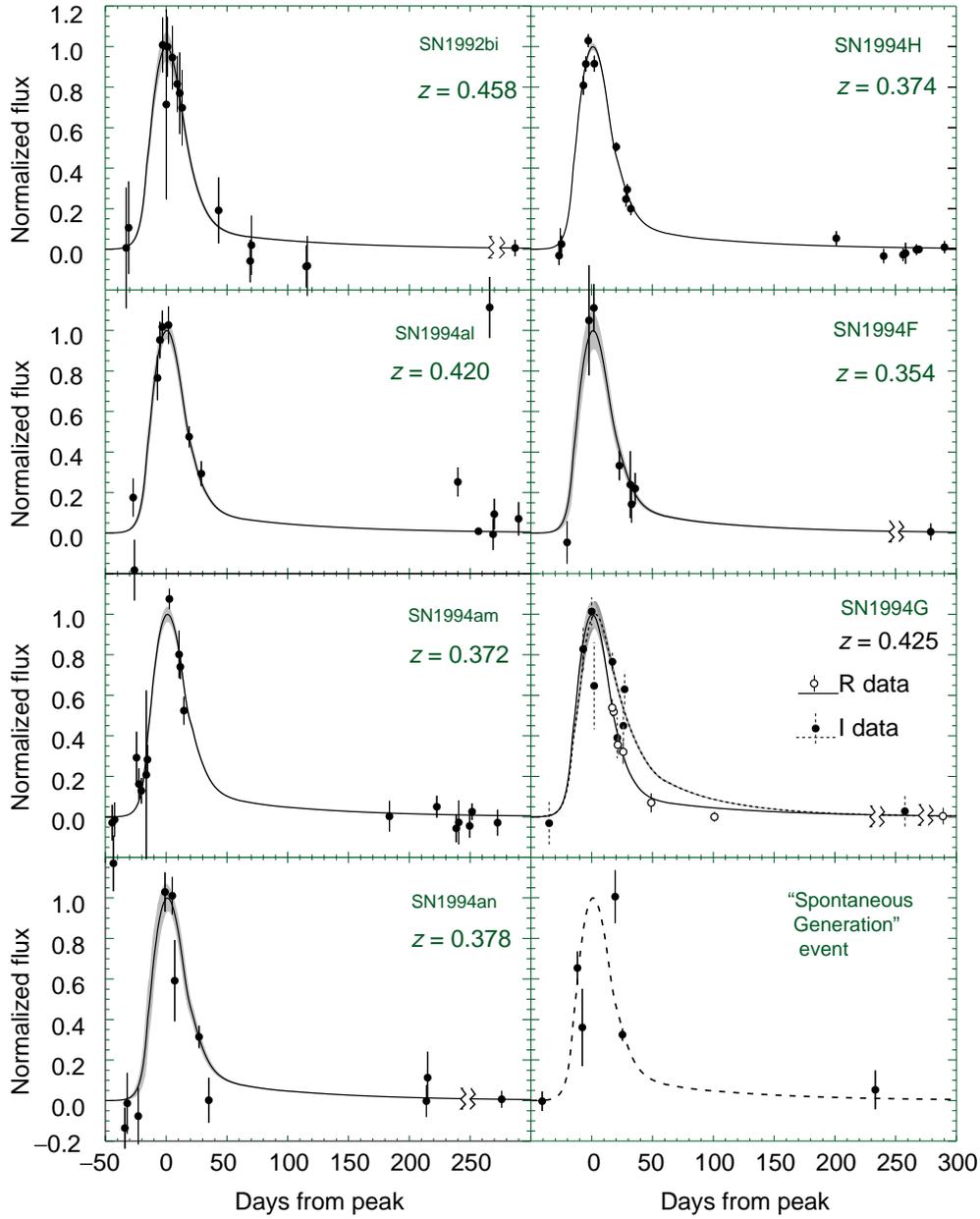,height=6.5in}
\caption[figurecaption]{
  Preliminary $R$-band light curves for the first seven high-redshift supernovae.  Shading behind the
solid curves represent the 1-$\sigma$ bounds on the best fit Leibundgut template light curve.
An $I$-band light curve is also shown for SN 1994G; other photometry points in $I$ and $B$
for these supernovae are not shown on this plot.
The lower
right panel shows the light curve
of a ``spontaneous generation'' event, a transient point-source of light on a region of an
image where no host galaxy is visible; the follow-up 
photometry was not as extensive for this event. }
  \label{lightcurves}
\end{figure}

In short, this search technique allows ``batches'' of 
pre-maximum-light supernova discoveries to be {\em scheduled} 
just before new moon, the ideal time to begin follow-up 
spectroscopy and photometry.  This follow-up can now be 
scheduled as well, on the largest telescopes.

\section{Supernova Discoveries and Follow Up}

As of this meeting in June 1995, we had used this 
``batch'' search technique several times at the Isaac 
Newton Telescope and the Kitt Peak 4-meter telescope to 
discover 9 supernovae.  Seven of these were found within the 
standard 2.5 -- 3 week search interval and were followed 
with photometry and spectroscopy (Perlmutter {\it et al.} 
\nocite{sn1992bi}\nocite{sne1994}\nocite{pe:foursn}
\nocite{pe:elevensn}1994, 1995a, 1995b).  (Since these were the 
demonstration runs of the project, not all of the follow up 
was scheduled.)  We observed light curves for all of the 
supernovae in at least one filter (usually R band), and 
spectra for all of the host galaxies and three of the 
supernovae.

[Since the Aiguablava meeting, we have now completed another 
search run using the CTIO 4-meter telescope.  This yielded 11 
supernovae at redshifts between 0.39 and 0.65 (Perlmutter 
{\it et al.} 1995c).]

The results of the batch search strategy can be seen in 
Figure 2, which shows a preliminary analysis of the 
$R$-band light 
curves as of June 1995.  For each of the supernovae, there 
is the 2.5 -- 3 week gap in the observations leading up to 
the discovery just before maximum light.  (There is also a 
gap during the following full moon.)  Figure 3a shows the 
distribution of discovery dates spread over the week before 
maximum light.  In the case of SN 1994am, the discovery 
was just after maximum light, but in this case the very 
first observations provide data points before maximum 
light.

\begin{figure}
\psfig{figure=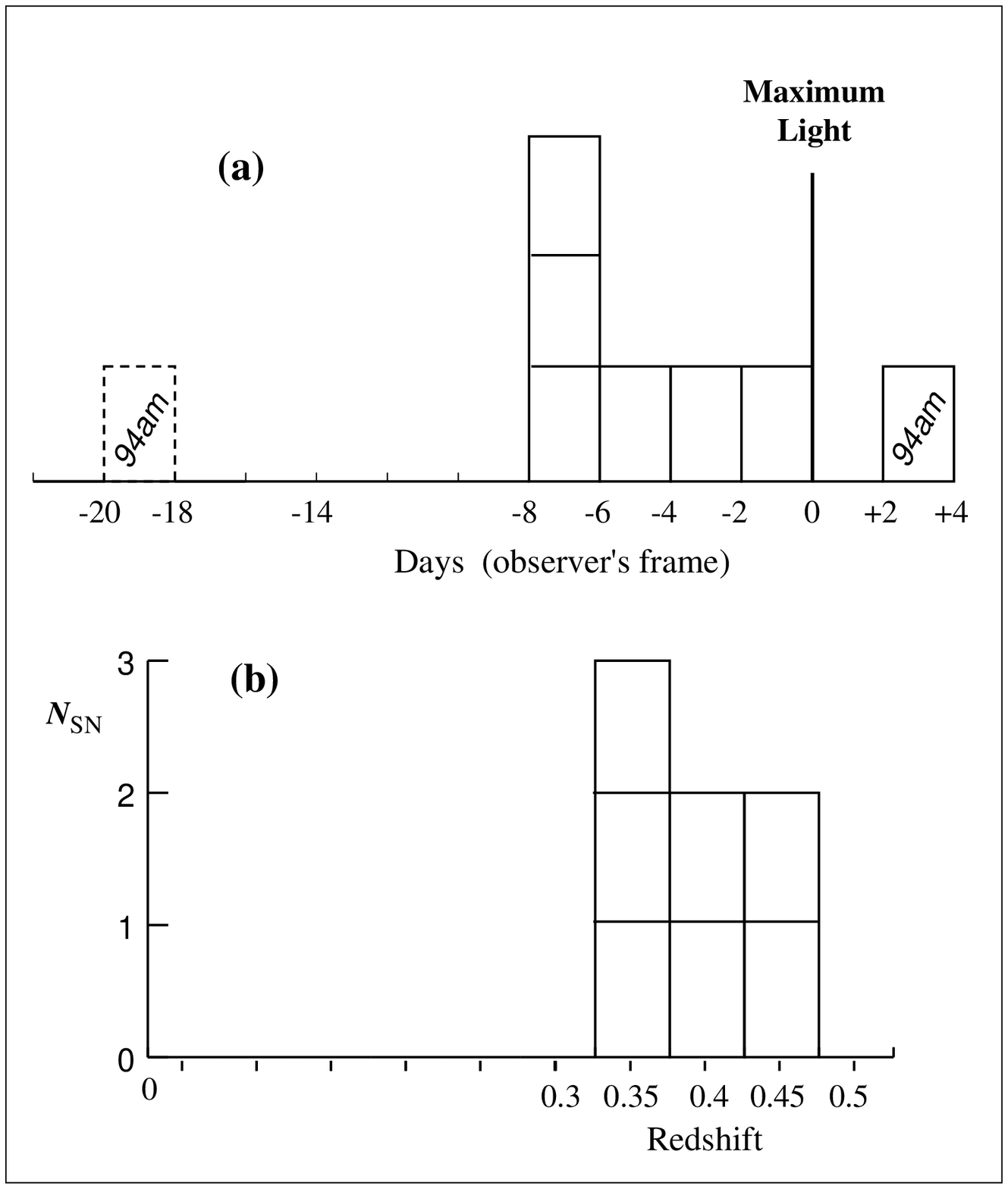,height=4in}
\caption[figurecaption]{
 (a) Distribution of discovery dates, plotted as 
number of days before maximum light.  Note that for SN 1994am the discovery 
was just after maximum light, but  the 
first observations, about three weeks earlier, provide a data point well before maximum 
light, shown in dotted outline.
(b) Distribution of redshifts.}
  \label{histo}
\end{figure}

Whenever possible, photometry was also obtained in other
bands in addition to $R$.   The light curve of SN 1994G shows
an example with more extensive coverage in both $I$ and $R$.
Several of the supernovae were observed in the $B$ band, 
although these points are not plotted in Figure 2.
$B$-band photometry near maximum light is
particularly important, because it gives an indication of 
rest-frame ultraviolet flux, as discussed below.

The redshift distribution of the supernova discoveries 
depends, of course, on the telescope aperture
and exposure 
times used for the search.  Most of these seven supernovae 
were discovered with data that was optimal for detection of 
supernovae in the redshift range $z$ = 0.35 -- 0.45, as 
discussed at this meeting in more detail by Reynald Pain and 
Isobel Hook (Pain {\it et al.}, 1996).  Figure 3b shows the actual 
redshift distribution, which matches the expected 
distribution quite well.  Our more recent searches use 
observations that are over a magnitude deeper and hence 
favor redshifts in the range 0.4 -- 0.6.

Two of the three supernova spectra observed, for SN 1994G 
(by Challis, Riess, and Kirshner) and SN 1994an, match well 
feature-for-feature to a nearby SN Ia spectrum observed on 
the corresponding  number of days (supernova rest frame) after 
maximum light, after redshifting the comparison spectrum to 
the redshift of the distant supernova's host galaxy.  The 
one spectrum that extends far enough into the red shows the  
Si II feature of an SN Ia.  The third supernova spectrum, for SN 
1994F, was observed during commissioning of the LRIS 
spectrograph (by J.B. Oke, J. Cohen, and T. Bida)  on the 
Keck telescope and is poorly calibrated.  It nevertheless 
shows peaks and troughs matching a redshifted SN Ia.  (The 
three spectra will be presented in forthcoming 
publications.)

Although no supernova spectrum was obtained for SN 1994am, its
host galaxy spectrum exhibits a large 4000-angstrom break
and lacks strong emission lines, indicating an elliptical galaxy.  
This supernova can therefore be taken to 
be an SN Ia as well.  Given that the lightcurve shapes  are not
consistent with SNe IIP,
and the higher probability of finding SNe Ia given their
brightness relative to the typical SNe II's, Ib's, and Ic's,
it is likely that all seven of the supernovae considered here
are SNe Ia (see Perlmutter {\em et al.} 1995a for further discussion
of SN 1992bi).

\section{Photometry Analysis}

The data analysis involves several stages, first to 
reduce the
observed image data to individual photometry points and then 
to compare these points with nearby SN Ia light curves to 
determine the luminosity distance and hence $q_0$.  The 
supernova light in each image must be measured and the 
underlying host galaxy light subtracted off.  At these 
redshifts the supernova's seeing disk usually covers a 
significant amount of the galaxy, so it is important 
to do this step correctly.  Each image has different seeing, and 
often a different telescope's point-spread-function (PSF) 
varies both spatially over the field and temporally 
over the course of an 
observing night.  The amount of galaxy light that must be 
subtracted therefore varies from image to image.

To ensure that this step does not introduce spurious 
fluctuations in the photometry, we have developed various 
analysis tests and controls to monitor the stability of the 
measurement.  For example, we repeat the analysis used at 
the location of the supernova at many other locations on 
nearby galaxies similar to the supernova's host galaxy.  
These locations should show flat ``light curves,"  since 
they have no supernova.  If the brightness fluctuates more 
than expected from the sky noise, we have evidence for a 
mismatch of galaxy light from image to image, due for 
example to a poor measurement of the seeing or local PSF.

These neighboring-galaxy locations also help monitor the accuracy 
of the color corrections that account for slightly different 
filters or CCD responses between telescopes and cameras.  As 
another check, we place ``fake'' supernovae on the images, 
and follow the analysis chain all the way through to the 
light curves to see that the analysis does not distort the photometry.

\section{Light Curve Analysis}

To compare these high-redshift supernova
photometry points 
to nearby SN Ia light curves, it is necessary to 
calculate the $K$ correction that accounts for the 
redshifting of the light observed in a given filter.  The 
standard $K$ corrections give the magnitude difference between 
the light emitted in a given filter band and the light 
observed after redshifting in that same band.  Since at 
redshifts of order $z = 0.45$ the light received in the $R$ 
band corresponds approximately to the light emitted in the 
$B$ band, we calculate a generalization of the $K$ 
correction, $K_{RB}$, that gives the magnitude difference 
between the rest frame $B$ magnitude and the observed $R$ 
magnitude.  This is described in more detail at this meeting 
by Kim {\it et al.} (1996\nocite{ki:aigua}), and Kim, Goobar, \& Perlmutter 
(1996\nocite{ki:Kcorr}).

In the past year, the case has become quite strong that SNe 
Ia are a family of very similar events, not all identical.  
Hamuy {\it et al.} (1995) \nocite{ha:hubble}
and Riess, Press, and Kirshner (1995) 
\nocite{re:lcs}
present evidence indicating that this family can be 
described by a single parameter, essentially representing 
the shape or width of the light curve, and that this 
parameter is tightly correlated with the absolute magnitude 
at maximum.  The broad, slow-light-curve supernovae appear 
somewhat brighter, while the narrow, fast-light-curve 
supernovae are somewhat fainter.

Hamuy {\it et al.} used 
Phillips' (1993\nocite{ph:delta}) characterizations of this 
light curve width, $\Delta m_{15}$, the magnitude drop in 
the first 15 days past maximum, and fit their data to 
template light curves from supernovae representing various 
$\Delta m_{15}$ values.  Riess {\it et al.} added and subtracted 
different amounts of a ``correction template'' to a 
Leibundgut ``normal'' template (Leibundgut {\it et al.} 1991 and
references therein\nocite{le:sup}) to represent this same light 
curve variation.   

For our preliminary analysis, we have chosen a third 
alternative approach, to stretch or compress the time axis 
of the Leibundgut template by a ``stretch factor'' $s$.  We 
find that this simple parameterization gives variations on 
the light curve that fit the variety of SNe Ia and can be 
translated to Phillips' $\Delta m_{15}$ via the formula 
$\Delta m_{15} = 1.7 /s - 0.6$.  Calibrating nearby 
supernovae with this $s$-factor, we find the width-brightness 
relation can be characterized by a magnitude correction of 
$\Delta {\rm mag} = 2.35(1-s^{-1})$, which closely matches Hamuy 
{\it et al}'s $\Delta m_{15}$ correction.

At this meeting, and since, a number of other supernova observables 
have been discussed as indicators of a given supernova's 
brightness within the SN Ia family.  Branch 
et al. (1996)\nocite{br:aigua} have 
discussed the correlation with $B-V$ and $U-B$ color, while 
Nugent {\it et al.} (1996) \nocite{nu:seq}
has pointed out indicative spectral 
features, suggesting that temperature may provide a single 
theoretical parameter that characterizes the variations 
within the SN Ia family.  These recent developments suggest 
that our final reduced ``data product'' must include not 
only the $K$-corrected magnitude at peak, but also any 
available information on light-curve width, colors, and 
spectral features that might locate the supernova within the 
family of fast-and-faint to broad-and-bright SNe Ia.

\section{Preliminary Scientific Results}

Having gathered and reduced these data, what scientific 
results can we glean from them?  Other talks 
at this meeting from
the Supernova Cosmology 
Project group address the uses of this data to determine 
limits on the spatial variability of the Hubble constant 
(Kim {\it et al.} 1996), to demonstrate the time dilation of a 
``standard clock'' at cosmological distances (Goldhaber {\it et 
al.} 1996)\nocite{go:aigua}, and to provide a first direct measurement of the 
SN Ia rate at $z \sim 0.4$ (Pain {\it et al.} 1996)\nocite{pa:aigua}.  
We will here 
discuss the homogeneity and width-brightness relation of SNe 
Ia at $z \sim 0.4$ and implications for measuring $q_0$.

Note that all the results should be understood as preliminary,
since we are now concluding the final stages of calibration and
cross-checking the analysis.

To use high-redshift SNe Ia as distance indicators, we must first ask if the
width-brightness relation holds at $z \sim 0.4$.  In Figure 4, 
we plot the stretch
factor---or equivalently $\Delta m_{15}$---versus the $K$-corrected
and Galactic-extinction-corrected
absolute magnitude (for a given $q_0$ and $H_0$), and we 
find a correlation that
is consistent with that found for nearby 
supernovae (shown as a solid line).  A different
choice of $q_0$ would only change the scale on the magnitude axis (and
slightly jitter the magnitude values, since the supernovae are not all at exactly
the same redshift), but this would not affect the agreement between
the near and distant width-brightness relation.

\begin{figure}
\psfig{figure=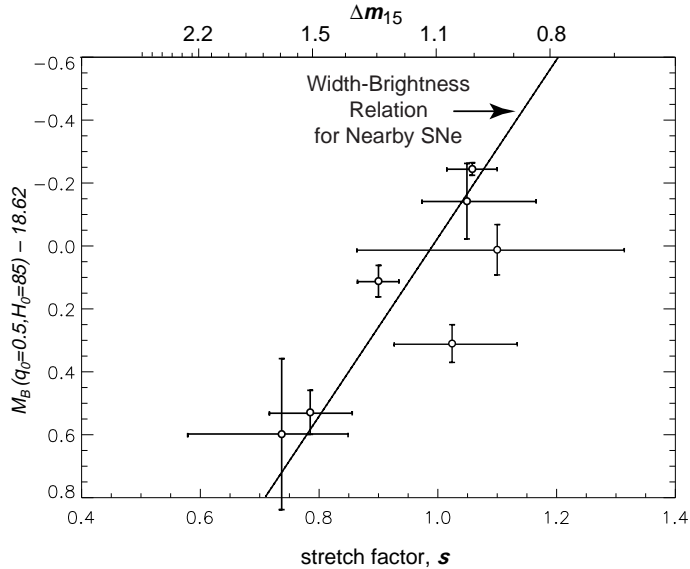,height=3in}
\caption[figurecaption]{
 K-corrected absolute B magnitude (for $q_0=0.5$ and 
$H_0 = 85$ m$\,$s$^{-1}$Mpc$^{-1}$) versus
stretch factor, $s$, representing the width of the SN Ia template light curve
that best fits the high-redshift supernovae.  The solid line shows
the relation between $M_B$ and $s$ found for nearby SNe Ia.   }
  \label{width}
\end{figure}

Using the width-brightness relation as a magnitude 
``correction,'' established using only nearby SNe Ia, 
tightens the dispersion of the high redshift supernova 
$K$-corrected peak magnitudes from 
$\sigma_{\rm raw} = 0.32$ mag 
to $\sigma_{\rm corrected} = 0.21$ mag.  These dispersions 
are quite comparable to those found for nearby SNe Ia before 
and after correcting for the width-brightness relation.  

When we turn to the other indicators of SN Ia family status, 
although the analysis is still incomplete, we find one good 
cross-check for the brightest of  our high redshift 
supernovae,  SN 1994H.  For this supernova, a $B-R$ color 
was measured at peak, which after $K$-correction is an 
indicator of  $U-B$ color in the supernova rest frame.  The
$B-R$ color indicates that SN 1994H is 
very similar to SN 1991T, which had a broad and bright light 
curve.  SN 1991T had an $s$-factor of $\sim$1.08, and our 
preliminary value for SN 1994H is $s = 1.06 \pm 0.08$, 
broader and brighter than any of the other 
high redshift supernovae of this sample with well-measured
lightcurve widths.  Ideally, these 
cross checks will be available for all future high-redshift 
supernovae.  This will be particularly important for cases 
for which the light curves are not complete enough for an 
accurate measurement of the lightcurve width.

\begin{figure}
\psfig{figure=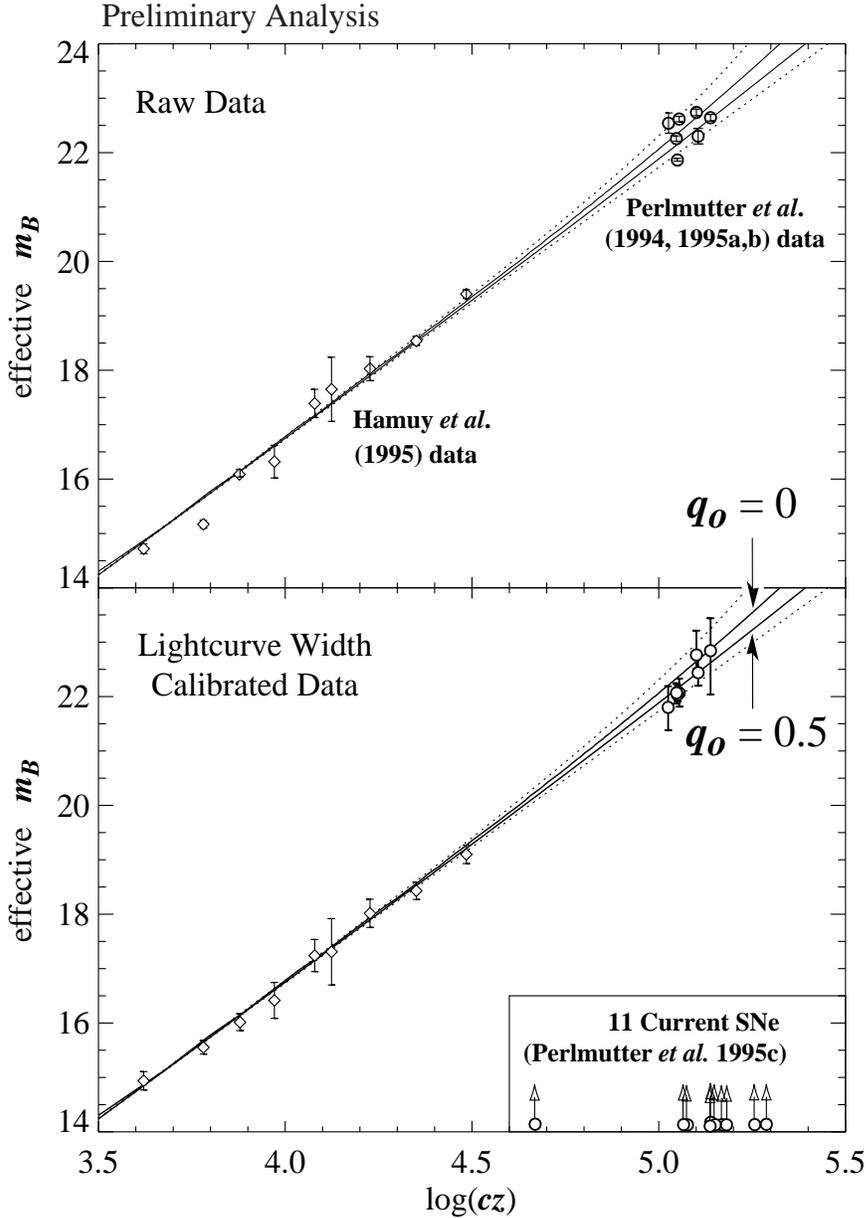,height=6.4in}
\caption[figurecaption]{
  Hubble diagrams for first seven supernovae (preliminary analysis). 
 Upper panel: ``raw'' apparent
magnitude measurements, after $K$ correction from $R$-band (observed)
to $B$-band (supernova rest frame), versus redshift.  The upper dotted curve
is calculated for $q_0 =-0.5$, the two solid curves are for $q_0=0$ and $q_0=0.5$,
and the lower dotted curve is for $q_0=1$.  Lower panel:  apparent
magnitude measurements after ``correction'' for width-brightness 
relation.  Note that the three points with the smallest error bars lie
on top of each other, in this plot.  
Inset:  Points represent redshifts for the most recent
11 supernova discoveries that are still being followed.}
  \label{hubble}
\end{figure}

For our preliminary estimate of $q_0$, we have used a few 
approaches.  The first is to apply the width-brightness 
correction to all seven of the high-redshift supernovae, on 
the assumption that they are all SNe Ia with
negligible extinction and that the zero
point of the width-brightness 
relation---the intercept of Figure 4---has not evolved 
in the $\sim$4 billion 
years back to $z \sim 0.4$.  Given the range of host galaxy ages
for the nearby supernovae used to derive this relation,  it is unlikely
that it would have a very different intercept (but the same slope) 
at $z \sim 0.4$.

Figure 5 shows the results of this analysis plotted on a Hubble
diagram, together with the relatively nearby SNe Ia of
Hamuy {\it et al.} (1995).  Comparing the upper and lower
panels, it is clear that the scatter about the Hubble line 
decreases 
for both near and high-redshift supernovae 
after correcting for the width-brightness relation, although the
error bars increase in the cases for which the light-curve width
($s$-factor or $\Delta m_{15}$) is poorly constrained by the photometry
data.

The three supernovae with the smallest error bars, after correction for
the width-brightness relation, are all near the redshift $z=0.37$ and
their data points in Figure 5, lower panel, lie on top of each other.  
They favor a relatively high value for $q_0$.  Note that before the
width-brightness correction (Figure 5, upper panel) their data points
spread from $q_0 \approx -0.5$ (upper dotted line) to $q_0 > 1$
(lower dotted line); generally, the correction
has brightened as many points as it has dimmed.

As an alternative approach to the $q_0$ measurement, we have also 
determined the value found for SN 1994G alone.  In the case of this supernova
we can avoid assumptions, because there is a strong spectrum for type 
identification, and its $B-R$ and $R-I$ colors are consistent with a ``normal''
SN Ia with no significant reddening. 
In particular, the (observed) $R-I$ color passes the (rest-frame)
$B-V$ cut proposed in Vaughan {\em et al.} \nocite{va:lmfn}(1995) to test for normal,
unreddened SNe Ia.
This is also consistent with its stretch factor, which is within error 
consistent with an $s=1$ standard Leibundgut template.
For SN 1994G,
we find a preliminary value of $q_0 = 0.8 \pm 0.35 \pm 0.3$, where the
second error is a preliminary bound on systematic error including
a $-0.1$ estimate for possible Malmquist bias discussed below.  This
is consistent with the width-brightness-corrected result
for the whole sample of seven supernovae.  (We re-emphasize that these
values are currently undergoing their final calibrations and cross-checks,
although they are unlikely to change significantly.)

\section{Discussion and Conclusion}
Given the error bars, our current measurements of $q_0$ do not yet clearly
distinguish between an empty $q_0=0$ and closed $q_0 > 0.5$ universe.  
The data do, however, indicate that a decelerating $q_0 \ge 0$ is a 
better fit than an accelerating $q_0 < 0$ universe.  This is an important
conclusion since it limits the possibility that $q_0 = \Omega_0 /2 - 
\Omega_\Lambda$ is dominated by the cosmological constant $\Lambda$
(where $\Omega_\Lambda$ is the normalized cosmological 
constant  $\Lambda (3 H_0)^ {-2}$).
In an accelerating universe, high
values for the Hubble constant do not conflict with the ages
of the oldest stars, because the
universe was expanding more slowly in the past.
However, in a ``flat'' universe, in which $\Omega_0 + \Omega_\Lambda
 = 1$, a  cosmological constant $ \Omega_\Lambda \ge 0.5$ would give
an acceleration $q_0 \le -0.25$, a poor fit to our data. 

Note that extinction of the distant supernovae would make this evidence of
deceleration even
stronger, as would the possibility of inhomogeneous matter distribution
in the universe as described
by Kantowski, Vaughan, \& Branch \nocite{ka:clump}(1995), since both
of these would lead to underestimates of $q_0$ if not taken into 
account.  (Our future analyses will
bound or measure extinction using multiple colors to differentiate reddening
from intrinsic color differences within the SN Ia family.  Riess
{\em et al.}\nocite{ri:aigua} (1996) discussed this approach at this conference.) 

 It will
be important to compare the $q_0$ value from our most distant supernovae to the
value from our closest high-redshift supernovae to look for Malmquist bias.
An estimate of the size of this bias using our studies of our detection 
efficiency as a function of magnitude
suggests that a 0.2 mag intrinsic dispersion in calibrated SN Ia
magnitudes would lead to an overestimate of  0.1 in $q_0$, if not accounted for 
(Pain {\em et al.} 1996, and see also Schmidt {\em et al.} 1996).

We will complete the observations and analyses of the next 11 supernovae soon, 
and be able to greatly strengthen the certainty of these conclusions.   The
new data set is expected to have
significantly smaller error bars, particularly after calibration
of the width-brightness relation, and it should be possible to distinguish
the empty, $q_0 = 0$, and closed, $q_0 > 0.5$, universes.  
The redshift of these supernovae are 
plotted in the inset of Figure 5.  Note that the highest redshift supernova,
at $z \sim 0.65$, would be over 0.35 magnitudes brighter in a closed universe
than in an empty universe.

For the future, it should be 
possible to measure both $\Omega_0$ {\em and}
$\Lambda$, even if the universe is not assumed to be flat.  The apparent
magnitude of a ``calibrated candle'' depends on both  $\Omega_0$ 
and $\Lambda$, but with different powers of $z$.  The discovery of still
higher redshift supernovae, at $z \sim 1$, can therefore locate our
universe in the $\Omega_0$ vs. $\Lambda$ parameter plane, as shown in
Figure 6.  The feasibility of this measurement is discussed in 
Goobar and Perlmutter \nocite{go:lambda}(1995).

\begin{figure}
\psfig{figure=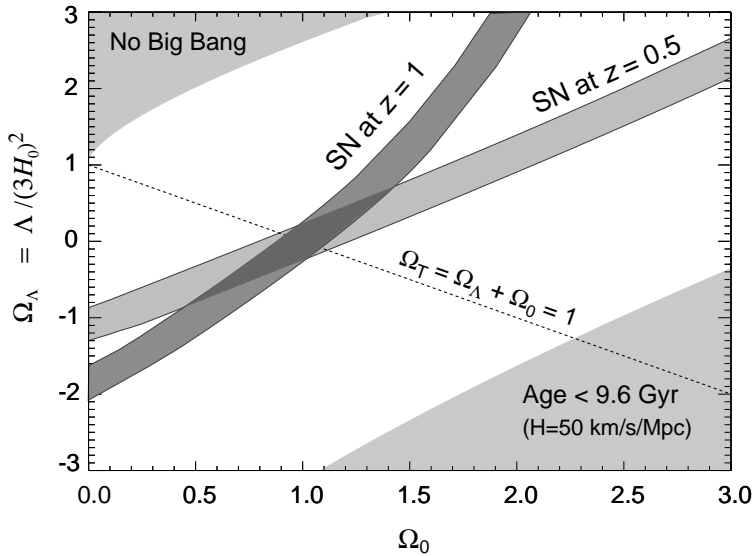,height=3in}
\caption[figurecaption]{
 Theoretical bands of allowed parameter space in the $\Omega_{\Lambda}$ versus  
$\Omega_0$ plane, for the hypothetical 
example of two SNe Ia discovered at $z=0.5$
and at $z =1$ in a universe with  $\Omega_{\Lambda} = 0$ and  $\Omega_0$ =1
(from Goobar and Perlmutter 1995).  The width of the bands is due
to uncertainty in the apparent magnitude measurement of the supernovae.  
The intersection region would provide a measurement 
of both $\Lambda$ and $\Omega_0$.  The shaded corners
are ruled out by other observations, as indicated.}
  \label{lambda}
\end{figure}

\vspace{0.22in}

We have shown here that the scheduled discovery and follow up of batches of 
pre-maximum 
high-redshift supernovae can be accomplished routinely, that the 
comparison with nearby supernovae can be
accomplished with a generalized K-correction, and that a width-brightness
calibration can be applied to standardize the magnitudes.  
Ideally this will now become a standard
method in the field, and SNe Ia beyond $z = 0.35$ will become a 
well-studied distance indicator---as they already are for $z<0.1$---useful
for measuring the cosmological parameters.  The prospects for this look
good:  already several other supernova groups have now
started high-redshift searches.  Schmidt {\em et al} \nocite{sc:aigua}(1996) and
Della Valle {\em et al} \nocite{de:aigua}(1996), at this meeting discussed two of these.

The many presentations at this exciting meeting have discussed 
 recent advances in our empirical and theoretical understanding of Type Ia supernovae.  
The resulting supernova-based measurement of the cosmological parameters 
 is an impressive consequence of these
efforts of the entire supernova
research community.

\vspace{.25in}

The observations described in this paper were primarily
obtained as visiting/guest
astronomers at
the Isaac Newton and William Herschel Telescopes, operated by
the Royal Greenwich Observatory at the Spanish Observatorio del Roque
de los Muchachos of the Instituto de Astrofisica de Canarias;
the Kitt Peak National Observatory 4-meter and 2.1-meter telescopes and
Cerro Tololo Interamerican Observatory 4-meter telescope,
both operated by the National Optical Astronomy Observatory under contract to
the National Science Foundation; the Keck Ten-meter Telescope; and the
Siding Springs 2.3-meter Telescope of the Australian National University.  We thank the staff of these observatories for their excellent 
support.  Other observers contributed to this data as well; in particular,
we thank Marc Postman, Tod Lauer, William Oegerle, and John Hoessel for their
more extended participation in the observing.
This work was supported in part by the Physics Division, 
E. O. Lawrence Berkeley National Laboratory of
the U.~S. Department of Energy under Contract No. DE-AC03-76SF000098,
and by the National Science Foundation's
Center for Particle Astrophysics, University of California,
Berkeley under grant No. ADT-88909616.

\bibliography{saul}
\bibliographystyle{plain}
\end{document}